\documentclass[epjST]{svjour}
\usepackage{graphics}
\begin{document}
\title{Chimera States in a Two--Population Network of Coupled Pendulum--Like Elements}
\author{Tassos Bountis\inst{1}\fnmsep\thanks{\email{tassos50@otenet.gr}} \and Vasileios G. Kanas \inst{2} \and Johanne Hizanidis\inst{3}
\and Anastasios Bezerianos\inst{4}}

\institute{Department of Mathematics, University of Patras, Patras, Greece \and Department of Electrical and Computer Engineering, University of Patras, Patras, Greece
\and National Center for Scientific Research ``Demokritos'', Athens, Greece \and Cognitive Engineering Lab,Singapore Institute for Neuroengieering (SINAPSE), National University of Singapore, Singapore}
\abstract{
More than a decade ago, a surprising coexistence of synchronous and asynchronous behavior called the chimera state was discovered in networks of nonlocally coupled identical phase oscillators. In later years, chimeras were found to occur in a variety of theoretical and experimental studies of chemical and optical systems, as well as models of neuron dynamics. In this work, we study two coupled populations of pendulum-like elements represented by phase oscillators with a second derivative term multiplied by a mass parameter $m$ and treat the first order derivative terms as dissipation with parameter $\epsilon>0$. We first present numerical evidence showing that chimeras do exist in this system for small mass values $0<m<<1$. We then proceed to explain these states by reducing the coherent population to a single damped pendulum equation driven parametrically by oscillating averaged quantities related to the incoherent population.   
} 
\maketitle
\section{Introduction}
\label{intro}
About ten years ago, a novel dynamical phenomenon was discovered in populations of identical and symmetrically coupled oscillators: Under nonlocal coupling, which generally decays with distance, a surprising coexistence of synchronized and asynchronous subpopulations was revealed and given the name {\it chimera state} after the Greek mythological creature made up of different animals. 

Chimera states were first reported by Kuramoto and Battogtokh in a model of densely and uniformly distributed oscillators, described by the complex Ginzburg-Landau equation in one spatial dimension, with nonlocal coupling of exponential form \cite{KUR02}. This was followed by the work of Abrams and Strogatz \cite{ABR04}, who observed this phenomenon in a 1--dimensional ring continuum of phase oscillators assuming nonlocal coupling with a cosine kernel and coined the word ``chimera'' for it. The same authors also found chimera states in networks of identical, symmetrically coupled Kuramoto phase oscillators \cite{ABR08} by considering two subnetworks with all--to--all coupling both within and between subnetworks, assuming strong coupling within each population and weaker coupling between them. 

More recently, Laing and co--authors used the same model to demonstrate the presence of chimeras in coupled Stuart--Landau oscillators \cite{LAI10} and investigated the effect of random removal of network connections on the existence and stability of chimera states \cite{LAI12}. Chimeras have also been observed in many other systems, including coupled chaotic logistic maps and R\"ossler models \cite{OME11}. The first experimental evidence of such states was subsequently reported in populations of coupled chemical oscillators, as well as in optical coupled--map lattices realized by liquid--crystal light modulators \cite{TIN12,HAG12}.

Concerning the importance of chimera states in brain dynamics, it is believed that they could potentially explain the phenomenon of unihemispheric sleep observed in many birds and dolphins which sleep with one eye open, suggesting that one hemisphere of the brain is synchronous the other being asynchronous \cite{RAT00,MAT06}. For this reason it is particularly interesting that such states were recently observed in FitzHugh--Nagumo \cite{OME13} and Hindmarsh--Rose \cite{HIZ14} networks of coupled oscillators modeling neuron dynamics.

Synchronization in phase oscillator systems with an inertial term involving a second order derivative multiplied by a mass parameter $m>0$ has been theoretically studied by some researchers (see e.~g. the review \cite{ACE05}). In some of these works an external periodic driving is included and a Fokker--Planck analysis is performed relating critical behavior and synchronization transitions to those of the corresponding Kuramoto model in the limit $m\rightarrow 0$. This raises the interesting question whether chimera states can actually be observed in physical systems of this type.

Recently, this question was answered affirmatively in an experiment involving two subpopulations of identical mechanical metronomes, whose inter--population coupling is weaker than the coupling within each subpopulation \cite{MAR13}. In this setting, chimeras were shown to emerge in a thin region of parameter space as a competition between two fundamental synchronization states. A variety of complex chimera--like states was observed and a mathematical model was proposed with variable damping and all oscillator masses equal to unity.

In this paper we demonstrate the existence of chimera states in two non--locally coupled populations of pendulum--like elements. Our model extends similar systems of phase oscillators studied in \cite{LAI10,LAI09,LAI12a} by including an inertial second derivative term and multiplying the first derivatives by a damping parameter $\epsilon$. As in these earlier studies, we assume constant coupling within each population, while the coupling strength between the two populations is weaker.

Our first result is that with $\epsilon =1$ chimera states are numerically observed for sufficiently small values of the mass parameter $0<m<<1$. In fact, these states occur within narrow intervals of mass values, with synchronization observed on either the left or right half of the total population. We then keep the mass fixed and gradually decrease the damping parameter $\epsilon>0$, until we reach a threshold value $\epsilon(m)$, below which chimeras are no longer observed.

Our results suggest that chimeras represent attracting states that cannot be continued down to the conservative (Hamiltonian) case $\epsilon=0$. We further support these findings by describing qualitatively our chimera states through a reduction of the synchronized population to a single damped pendulum driven parametrically by oscillating averaged quantities obtained from the asynchronous population of the network. 

\section{Two Coupled Populations of Pendulum--like Elements}
\label{sec:1}
Let us consider a two population network of non--identical phase oscillators in the form studied in \cite{LAI10,LAI12,LAI09,LAI12a} and extend it by introducing inertial terms proportional to a mass parameter $m>0$. We thus obtain a network of two populations of pendulum--like elements with all--to--all coupling within (and between) the populations, governed by the second order ordinary differential equations (ODEs):

\begin{equation}
m\frac{d^2\theta_{i}^{1}}{dt^2} + \epsilon\frac{d\theta_{i}^{1}}{dt}= \omega_i - d_1sin(\theta_{i}^{1}) + \frac{\mu}{N}\sum_{j=1}^{N} {\sin ({\theta_j^{1}-\theta_i^{1}-\alpha})} + \frac{\nu}{N}\sum_{j=1}^{N} {\sin ({\theta_j^{2}-\theta_i^{1}-\alpha})}
\label{eq:eq1}
\end{equation}
\begin{equation}
m\frac{d^2\theta_{i}^{2}}{dt^2} + \epsilon\frac{d\theta_{i}^{2}}{dt}= \omega_i - d_2sin(\theta_{i}^{2}) + \frac{\mu}{N}\sum_{j=1}^{N} {\sin ({\theta_j^{2}-\theta_i^{2}-\alpha})} + \frac{\nu}{N}\sum_{j=1}^{N} {\sin ({\theta_j^{1}-\theta_i^{2}-\alpha})}
\label{eq:eq2}
\end{equation}

where $i,j = 1,...N$, $N$ is the number of oscillators in each population (subnetwork). The two populations are labeled by the superscripts 1 and 2, while $\mu>0$ and $\nu>$ are fixed parameters representing the coupling strength within the same population and between the two populations respectively. The $\omega_i$ are taken from a Lorentzian distribution $g(\omega)$ \cite{LAI10}, while $\mu$ and $\nu$ satisfy $\mu + \nu = 1$, with $\mu > \nu$ as in \cite{LAI12}. The oscillators in population 1 are numbered $1$ to $500$, while those in population $2$ are numbered $501$ to $1000$.

Note that equations \ref{eq:eq1} and \ref{eq:eq2} describe the motion of two subnetworks of pendulum--like oscillators, with each pendulum  characterized by a mass $m$ and a damping parameter $\epsilon$. In what follows, we will first neglect gravity, setting $d_1=d_2=0$ in \ref{eq:eq1} and \ref{eq:eq2} and later allow these parameters to be non--zero. 

Now, as chimera states represent {\it attractors} of the dynamics, we do not expect them to exist in the conservative limit $\epsilon=0$. We, therefore, attempt to locate chimeras in two ways: (a) First we set $\epsilon=1$ and increase the value of $m$ gradually from 0 and (b) we fix $m$ at a value where a chimera is known to exist and investigate the range of decreasing $\epsilon<1$ values over which the phenomenon persists.

As in previous studies, we set $N = 500$, $\mu$ = 0.6, $\nu$ = 0.4, $\alpha$ = $\pi$/2 - 0.05 and use the Lorentzian distribution $g(\omega)=\frac{1}{\pi}\frac{\gamma}{\gamma^2+\omega^2}$ with $\gamma=0.001$. We also varied the $\alpha$ value and only detected chimeras very close to $\alpha = \pi/2 - 0.05$. Examining their dependence on the coupling coefficients we found that, with $\mu=0.6$, chimeras typically break down for inter--population coupling $0<\nu<0.3$. We note that for our numerical integration we have employed the fourth order Runge-Kutta method, while in all our simulations we use random initial conditions chosen uniformly within the interval $(0,2\pi)$ for both $\theta^1_i(0)$ and $\theta^2_i(0)$.

\subsection{Chimera states for $\epsilon=1$}
\label{sec:2}
\begin{figure}[tbp]
\centering
\resizebox{0.8\columnwidth}{!}{
\includegraphics{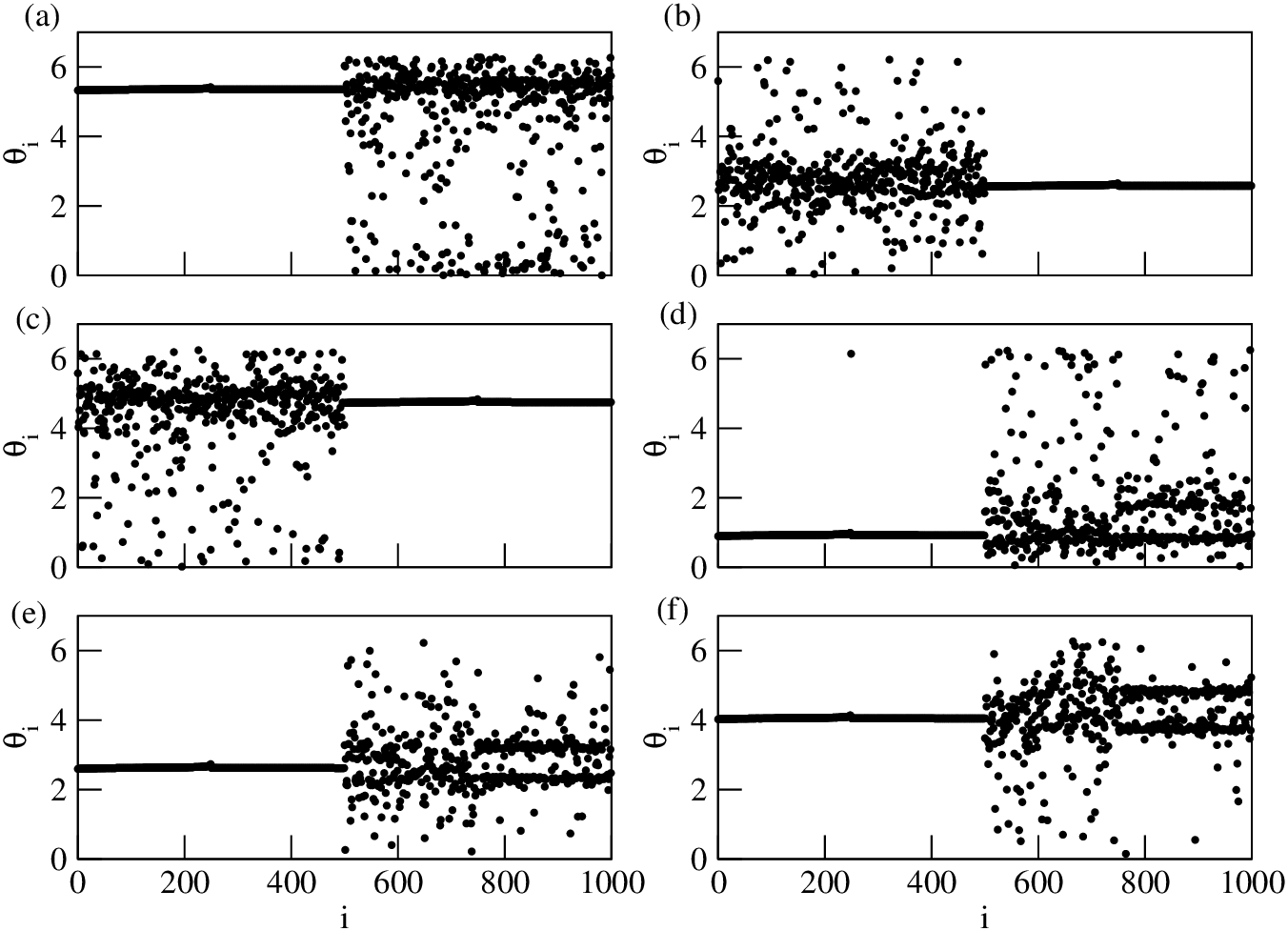}}
\caption{Snapshots of the variable $\theta_i$ for both populations at $t=3000$ 
for $\epsilon=1$ and increasing mass values: (a) $m=0.001$, (b) $m=0.0095$, (c) $m=0.011$, (d) $m=0.016$,
(e) $m=0.018$, and (f) $m=0.02$.}
\label{fig:fig1}
\end{figure}

Initially, we investigated the emergence of chimeras in parameter space by systematically increasing the value $m$ from 0, while the other parameters are kept constant and $\epsilon=1$. Theoretically, when $m=0$ our system reverts to the previously studied case. However, for $m\neq 0$, equations (\ref{eq:eq1}) and (\ref{eq:eq2}) constitute a second order system of ODEs and the Ott--Antonsen ansatz no longer applies. The results shown in Fig.~\ref{fig:fig1} demonstrate that increasing $m>0$ chimera states are found to exist over a narrow interval of small mass values.

Interestingly, these chimeras persist when we also take gravity into account modeled by the parameters $d_i, i=1,2$ in (\ref{eq:eq1}) and (\ref{eq:eq2}) greater than zero. Indeed, setting $m=0.011$, $\epsilon=1$ and the coupling parameters as above, we found that the chimera state of the $d_1=d_2=0$ case also exists for $d_1=d_2=0.001,0.01,0.1$. Thus, the elements we have considered in our equations may be thought of as pendulum--like.

\subsection{Chimera states for fixed $m$ and decreasing $\epsilon$}
\label{sec:3}
\begin{figure}[tbp]
\centering
\resizebox{0.8\columnwidth}{!}{
\includegraphics{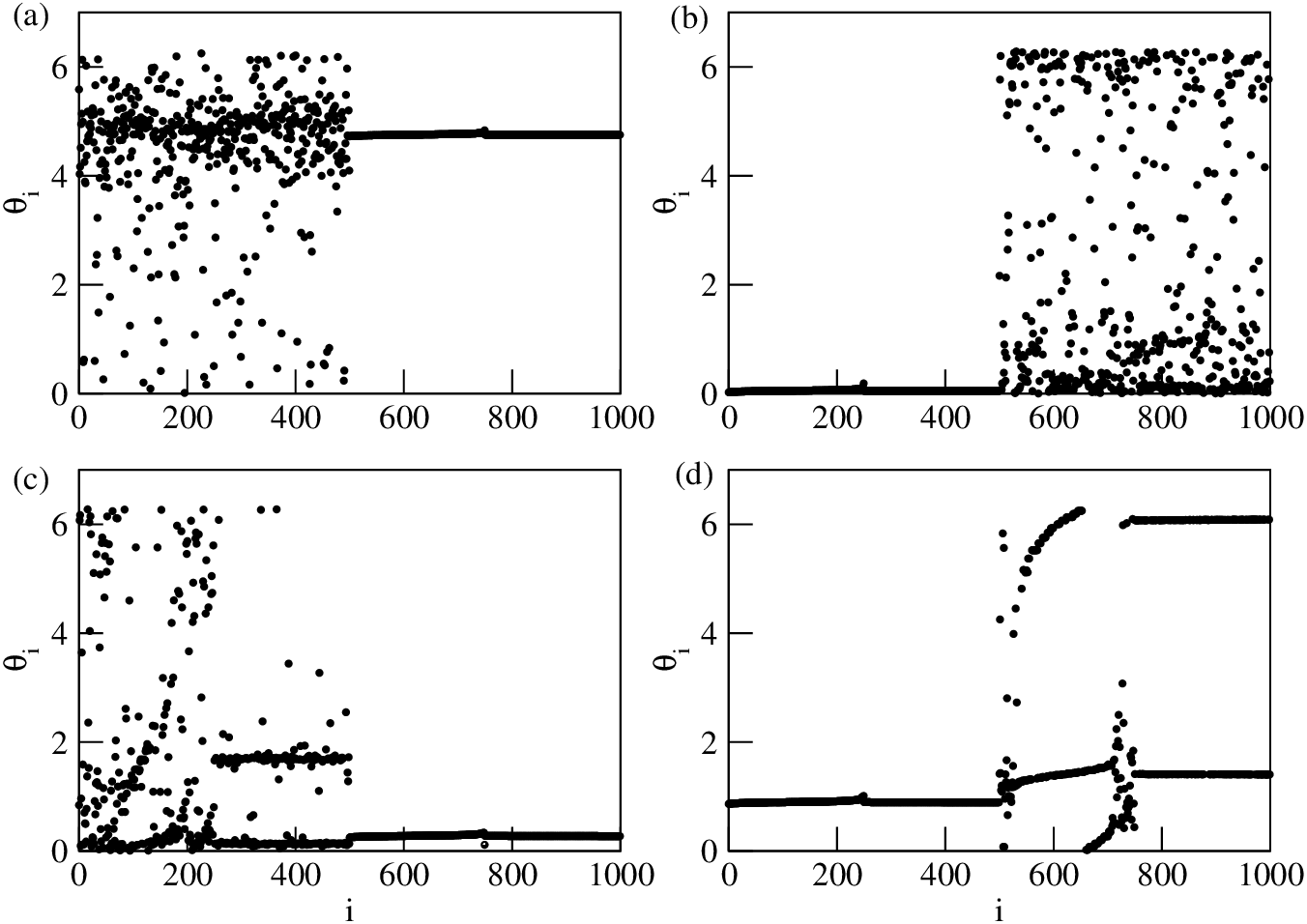}}
\caption{Snapshots of the variable $\theta_i$ at $t=3000$ 
for $m=0.011$ and decreasing damping rates: (a) $\epsilon=1$, (b) $\epsilon=0.9$, (c) $\epsilon=0.68$,
and (d) $\epsilon=0.56$.}
\label{fig:fig2}
\end{figure}

We now turn to examine the effect of decreasing the damping parameter $\epsilon$, for a mass value at which chimeras have been found to exist. For example, consider the case $m = 0.011$ of Fig.~\ref{fig:fig1} and start decreasing  $0 \leq \epsilon\leq 1$. What we observe in Fig.~\ref{fig:fig2} is that at these parameter values chimeras persist
down to approximately $\epsilon=0.61$, where intermittent cases of nearly full synchronization are evident. Below this threshold value chimeras are no longer observed and only patterns of nearly synchronized subgroups appear in the network.

\begin{figure}[tbp]
\centering
\vspace{20pt}%
\resizebox{0.5\columnwidth}{!}{
\includegraphics{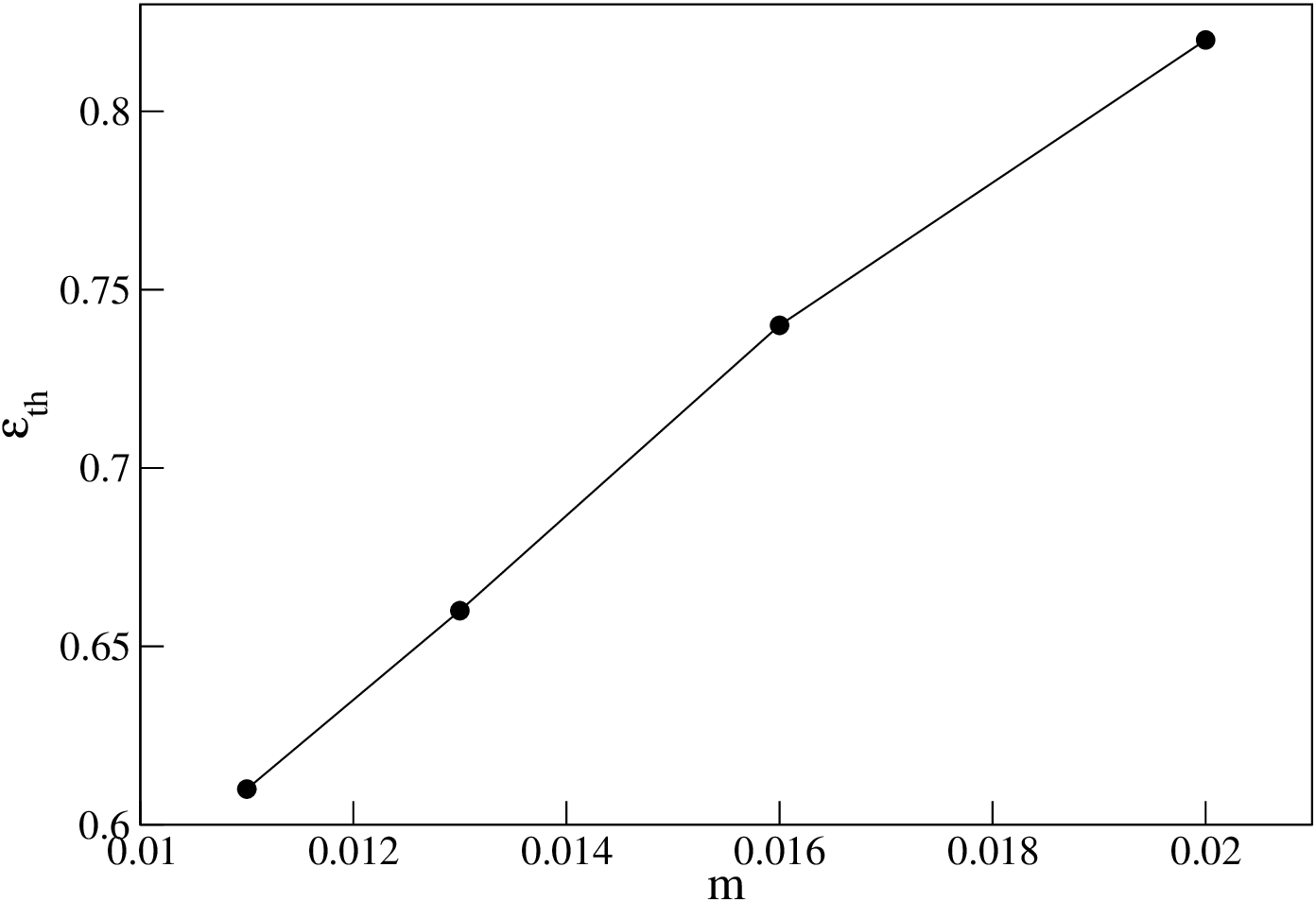}}
\caption{Threshold values $\epsilon_{th}=\epsilon(m)$, below which chimeras cease to exist. Note that $\epsilon(m)$ increases nearly linearly as a function of the mass $m$.}
\label{fig:fig3}
\end{figure}

We have thus repeated these calculations for different values of $m$ and estimated approximate $\epsilon(m)$ thresholds below which chimeras are seen to break down. Choosing $m=0.013, 0.016, 0.02$ and plotting the corresponding $\epsilon(m)=0.66, 0.74, 0.82$ values in the $(m,\epsilon_{th})$ plane of Fig.~\ref{fig:fig3}, we find that they seem to fall on a nearly straight line.

\section{Representing Chimeras by a Single Pendulum Equation}
\label{sec:4}
Let us now try to understand the chimera states we have discovered in a more qualitative way. For simplicity, we define the first population variables $\theta_{i}^{1}\equiv\phi_{i}$ and those of the second $\theta_{i}^{2}\equiv\theta_{i}$ in equations (\ref{eq:eq1}) and (\ref{eq:eq2}) and consider a typical chimera at $m=0.011$ and $\epsilon=1$ (see Fig.~\ref{fig:fig1} and Fig.~\ref{fig:fig2}), where the synchronized motion occurs in the second population. Moreover, let us also set $d_1=d_2=0$ and $\omega_{i}=\omega$ and equate the variables of the synchronized population to obtain the following single pendulum equation:

\begin{equation}
\label{reduce}
m\frac{d^2\theta}{dt^2} + \epsilon\frac{d\theta}{dt}= \omega - \mu \sin (\alpha) + \frac{\nu}{N}\sum_{j=1}^{N} {\sin ({\phi_j-\theta-\alpha})},
\end{equation}
where the first sum on the right has collapsed to a single term due to the fact that $\theta_{i}=\theta$, for all $i=1,...,500$. Clearly, the precise values of $\omega$, $\mu$ and $\alpha$ are not important since the first two terms on the right side of (\ref{reduce}) merely contribute to a constant average about which $\theta(t)$ oscillates.

Expanding now the sine in the sum appearing in (\ref{reduce}) produces two terms which may be expressed in terms of quantities averaged over the oscillations of the unsynchronized population as follows

\begin{equation}
\label{averaged}
s(t)=\frac{1}{N}\sum_{j=1}^{N} {\sin (\phi_j)},\,\,\,\ c(t)=\frac{1}{N}\sum_{j=1}^{N} {\cos (\phi_j)},
\end{equation}
so that Eq. (\ref{reduce}) finally becomes

\begin{equation}
\label{pendulum}
m\frac{d^2\theta}{dt^2} + \epsilon\frac{d\theta}{dt}= \omega - \mu \sin (\alpha) + \nu s(t)\cos (\theta) - \nu c(t)\sin (\theta),
\end{equation}
where we have also shifted the variable $\theta\rightarrow\theta - \alpha$ with no loss of generality. 

\begin{figure}[tbp]
\centering
\resizebox{\columnwidth}{!}{
\includegraphics{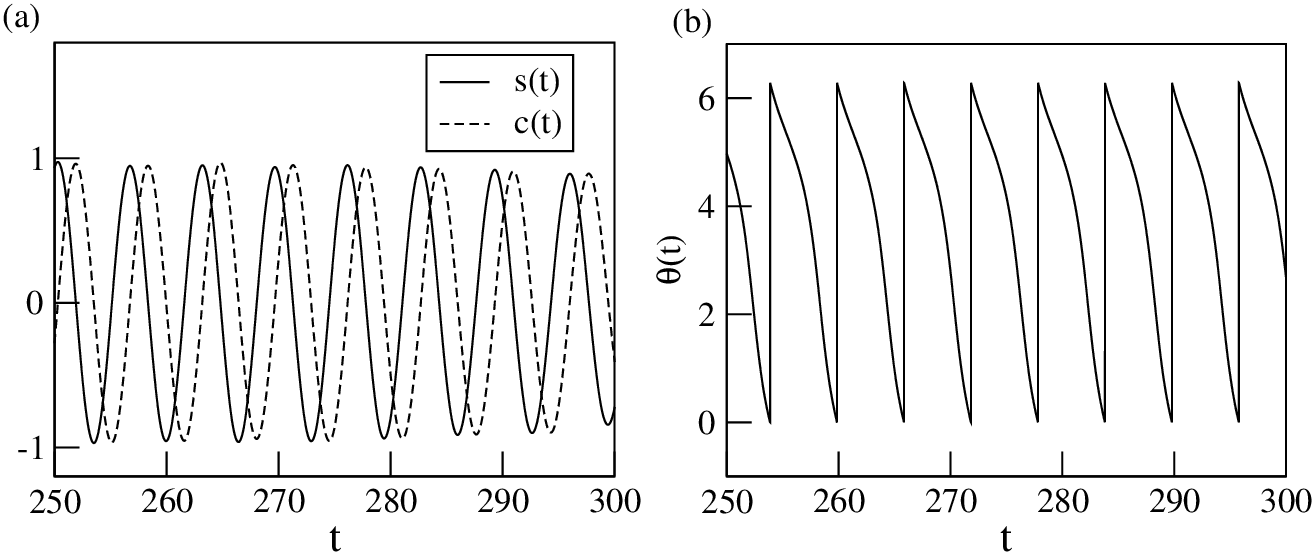}}
\caption{a) Oscillations of the averaged variables $s(t)$ (solid line) and $c(t)$ (dashed line) and b) the solution $\theta(t)$ of Eq. (\ref{pendulum}). The parameter values are $m=0.011$ and $\epsilon=1$.}
\label{fig:fig4}
\end{figure}

Our motivation for the above analysis stems from the numerical observation that the asynchronous elements oscillate on the average with the same period as the synchronous ones. This is clearly seen in Fig.~\ref{fig:fig4}, where the averaged variables $s(t)$ and $c(t)$ are plotted in panel (a), while in panel (b) the solution $\theta(t)$ of (\ref{pendulum}) is shown. Remarkably these quantities have the same period as the original variables, suggesting that the chimera state represents a phenomenon of entrainment due to periodic forcing. This is indeed verified by the fact that the oscillatory terms $s(t)$ and $c(t)$, which represent the effect of the incoherent population, enter parametrically in equation (\ref{pendulum}).

\begin{figure}[tbp]
\centering
\resizebox{0.5\columnwidth}{!}{

\includegraphics{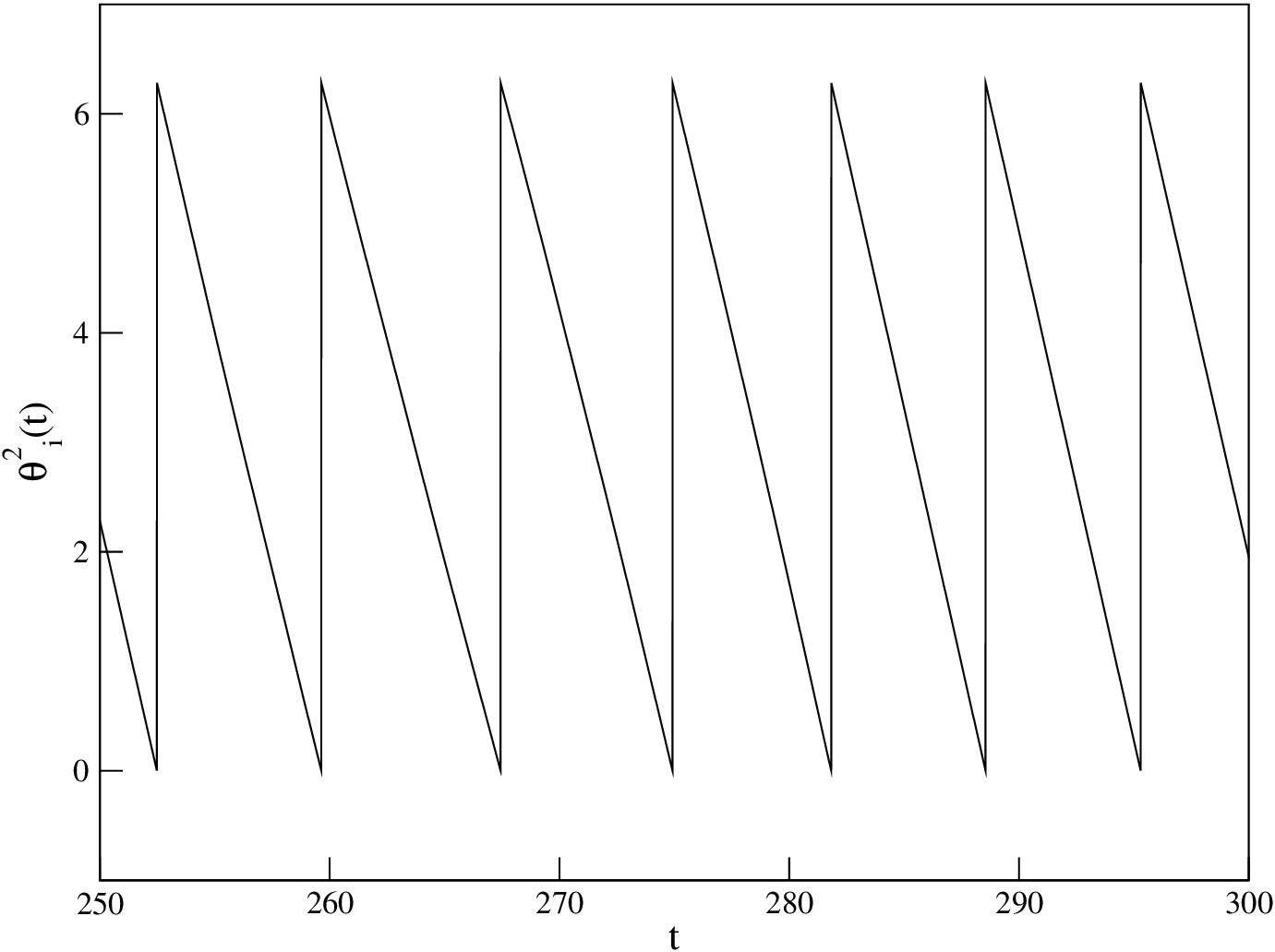}}

\caption{Oscillations of the synchronized variables $\theta_i(t)$ of Eq. (2). The parameter values were $m=0.011$ and $\epsilon=1$.}
\label{fig:fig5}
\end{figure}

It is remarkable that the amplitude and form of the $\theta(t)$ oscillations are clearly related to the corresponding quantities of the actual $\theta_i(t)$ oscillations shown in Fig.~\ref{fig:fig5}. Note, however, that there is a certain ``deformation'' in the oscillations of the reduced variable and a slightly smaller period than the one of the full system, which may be due to the imposed simplifications and need to be better understood. Nevertheless, we find it quite interesting that the above severe reduction of the problem to a single damped pendulum equation (\ref{pendulum}) has allowed us to propose a plausible qualitative explanation of such a complex phenomenon as the chimera state of a two-population network of pendulum--like oscillators.


\section{Discussion and Conclusions}
\label{sec:5}
Chimera states of coexisting synchronous and asynchronous populations can be observed in many networks of identical, symmetrically coupled phase oscillators, provided the coupling within each population is weaker than the inter--population coupling. Although chimeras have been theoretically and experimentally discovered in a variety of physical and chemical systems, their occurrence in more general physical and biological networks is just beginning to be explored.

Regarding mechanical systems, chimera states were recently found in an experiment involving two populations of identical metronomes, with all masses equal to unity and first order derivatives representing damping. To probe the generality of these results, we decided to investigate numerically the occurrence of chimeras in a network of phase oscillators, with all--to--all coupling, adding to the equations of motion a second derivative term and introducing a damping parameter $\epsilon$ before the first order derivatives. 

Keeping all other parameters constant and fixing $\epsilon =1$, we found that chimera states exist for relatively small values of the mass parameter $m$. On the other hand, when the mass of the oscillators is kept fixed and $\epsilon$ is decreased from $1$, a threshold $\epsilon(m)$ is discovered at which chimeras break down and are replaced by complicated states of synchronous and asynchronous motion for $0<\epsilon<\epsilon(m)$. Furthermore, as the value of $m$ grows, so does the value of $\epsilon(m)$, yielding points that appear to fall on a straight line. This dependence, however, requires a more detailed study, which we intend to carry out in a future publication.

Thus, our results show that in our model chimera states do not persist all the way to the conservative case $\epsilon=0$. This is perhaps not surprising since chimeras typically represent attracting states. On the other hand, an analysis of these chimeras along the steps outlined in \cite{LAI10} shows that their appearance is not the result of a saddle-node bifurcation \cite{LAI13}. It seems, therefore, that in our system of second order ODEs chimeras destabilize due to a more complicated type of bifurcation that clearly requires further investigation.

Finally, we carry out a reduction of our two population network to a single second order ODE describing the synchronized part of the chimera by a single angle variable $\theta(t)$. We are thus able to show that chimera states in our system can be qualitatively described by a single damped pendulum, which is parametrically driven by the periodic oscillations of averaged quantities related to the unsynchronized population of oscillators.

\section{Acknowledgments}
\label{sec:6}
One of the authors (T.~B.) gratefully acknowledges the hospitality of the New Zealand Institute of Advanced Study and during the period of February 20 -- April 15, 2013, when some of this work was carried out. He is thankful for many useful conversations he had during this visit on topics of chimeras and nonlinear dynamics with Professor Sergej Flach and Associate Prof. Carlo Laing of Massey University, New Zealand. This research has been co--financed by the European Union (European Social Fund – ESF) and Greek national funds through the Operational Program ``Education and Lifelong Learning'' of the National Strategic Reference Framework (NSRF) -- Research Funding Program: Thales. Investing in knowledge society through the European Social Fund.

%
%

%

\end{document}